\documentstyle[12pt]{article}

\begin{document}


\def\a{\alpha}
\def\b{\beta}
\def\c{\varepsilon}
\def\d{\delta}
\def\e{\epsilon}
\def\f{\phi}
\def\g{\gamma}
\def\h{\theta}
\def\k{\kappa}
\def\l{\lambda}
\def\m{\mu}
\def\n{\nu}
\def\p{\psi}
\def\q{\partial}
\def\r{\rho}
\def\s{\sigma}
\def\t{\tau}
\def\u{\upsilon}
\def\v{\varphi}
\def\w{\omega}
\def\x{\xi}
\def\y{\eta}
\def\z{\zeta}
\def\D{\Delta}
\def\G{\Gamma}
\def\H{\Theta}
\def\L{\Lambda}
\def\F{\Phi}
\def\P{\Psi}
\def\S{\Sigma}

\def\o{\over}
\def\beq{\begin{eqnarray}}
\def\eeq{\end{eqnarray}}
\newcommand{\gsim}{ \mathop{}_{\textstyle \sim}^{\textstyle >} }
\newcommand{\lsim}{ \mathop{}_{\textstyle \sim}^{\textstyle <} }
\newcommand{\vev}[1]{ \left\langle {#1} \right\rangle }
\newcommand{\bra}[1]{ \langle {#1} | }
\newcommand{\ket}[1]{ | {#1} \rangle }
\newcommand{\EV}{ {\rm eV} }
\newcommand{\KEV}{ {\rm keV} }
\newcommand{\MEV}{ {\rm MeV} }
\newcommand{\GEV}{ {\rm GeV} }
\newcommand{\TEV}{ {\rm TeV} }
\def\diag{\mathop{\rm diag}\nolimits}
\def\Spin{\mathop{\rm Spin}}
\def\SO{\mathop{\rm SO}}
\def\O{\mathop{\rm O}}
\def\SU{\mathop{\rm SU}}
\def\U{\mathop{\rm U}}
\def\Sp{\mathop{\rm Sp}}
\def\SL{\mathop{\rm SL}}
\def\tr{\mathop{\rm tr}}

\def\IJMP{Int.~J.~Mod.~Phys. }
\def\MPL{Mod.~Phys.~Lett. }
\def\NP{Nucl.~Phys. }
\def\PL{Phys.~Lett. }
\def\PR{Phys.~Rev. }
\def\PRL{Phys.~Rev.~Lett. }
\def\PTP{Prog.~Theor.~Phys. }
\def\ZP{Z.~Phys. }


\baselineskip 0.7cm

\begin{titlepage}
\begin{flushright}
UT-02-09
\end{flushright}

\vskip 1.35cm
\begin{center}
{\large \bf
Higher-Dimensional QCD without the Strong CP Problem
}
\vskip 1.2cm
Izawa K.-I.${}^{1,2}$, T.~Watari${}^{1}$ and T.~Yanagida${}^{1,2}$
\vskip 0.4cm

${}^1${\it Department of Physics, University of Tokyo,\\
     Tokyo 113-0033, Japan}

${}^2${\it Research Center for the Early Universe, University of Tokyo,\\
     Tokyo 113-0033, Japan}

\vskip 1.5cm

\abstract{
QCD in a five-dimensional sliced bulk with chiral extra-quarks
on the boundaries is generically free from the strong CP problem.
Accidental axial symmetry is naturally present
except for suppressed breaking interactions,
which plays a role of the Peccei-Quinn symmetry
to make the strong CP phase sufficiently small.
}
\end{center}
\end{titlepage}

\setcounter{page}{2}

\section{Introduction}

The standard model of elementary particles has
two apparent fine-tuning problems which
are hard to be undertaken directly by additional (gauge) symmetries:
the cosmological constant and the strong CP
\cite{KC}
problems.
The presence of extra dimensions might serve as
an alternative to symmetry 
in obtaining such fine-tuned parameters in a natural way.

In this paper, as a concrete example, we consider
QCD in a five-dimensional sliced bulk with chiral extra-quarks
on the boundaries
and point out that it is generically free from 
the strong CP problem.

For definiteness,
let us suppose that there is a pair of extra-quarks
in addition to the standard-model quarks:
a left-handed colored fermion $\psi_L$ and 
a right-handed one $\psi_R$.
We assume an extra-dimensional space
which separates them
from each other along the extra dimension.
If the distance between them is sufficiently large,
the theory possesses an axial U(1) symmetry
\begin{equation}
\psi_L \rightarrow e^{i \alpha} \psi_L, \qquad
\psi_R \rightarrow e^{-i \alpha} \psi_R 
\label{eq:chiral-transf}
\end{equation}
approximately,
whose breaking is suppressed
at a fundamental scale. 
This accidental global symmetry, which is actually broken by 
a QCD anomaly, plays a role of the Peccei-Quinn symmetry
\cite{PQ},
making the effective strong CP phase 
to be sufficiently small. 

The point is that the presence of such an approximate symmetry
is not an artificial requirement,
but a natural result stemming from the higher-dimensional space,
which might well be stable even against
possible quantum gravitational corrections.
Note that axion models in the four-dimensional spacetime suffer from
a fine-tuning problem:
operators that break the axial symmetry should be highly suppressed 
without any dynamical reasons.
The present model provides a solution to this problem.

\section{Bulk color gauge theory}

Let us consider a four-dimensional Minkowski spacetime $M_4$ 
along with one-dimensional extra-space $S^1$,
whose coordinate $y$ extends form $-l$ to $l$ (that is, two points
at $y=l$ and $y=-l$ are identified).
The SU(3)$_C$ gauge field is assumed to propagate on the whole
spacetime $M_4 \times S^1$.
The action of the five-dimensional gauge field is given by
\beq
 S_A =  
 \int \! d^4x \int_{-l}^l \! dy \ 
 {M_* \o 4 g^2(y)} \tr (F_{MN}F^{MN})
 + \int h(y) \tr \left( AF^2 - {1 \o 2}A^3F + {1 \o 10}A^5 \right),
 \label{eq:S1action}
\eeq
where $F_{MN} = \partial_M A_N - \partial_N A_M + [A_M , A_N ]$
$(M,N=0,\cdots,4; x^4 \equiv y)$,
$A=A_Mdx^M$, and $F=dA+A^2 = (1/2)F_{MN}dx^M dx^N$.
Here, $M_*$ is supposed to be a cutoff scale in the higher-dimensional 
gauge theory
and $g(y)$ and $h(y)$ are
gauge and Chern-Simons coupling functions, respectively.

Kaluza-Klein reduction to the four-dimensional spacetime, however, yields 
a massless color-octet scalar which is undesirable
in the low-energy spectrum.
Hence we consider an $S^1/{\bf Z}_2$ orbifold instead of the $S^1$.
The five-dimensional gauge field $A_M(x,y)$ is now under a constraint
\begin{equation}
 A_\mu(x,y)=A_\mu(x,-y), \quad 
 A_4(x,y)=-A_4(x,-y),
\end{equation}
where $\mu=0,\cdots,3$.
Then, we have only a vector field at low energies 
without the scalar field after the Kaluza-Klein reduction.
In order to define the theory on the orbifold consistently, 
the action eq.(\ref{eq:S1action}) on the $S^1$ should be 
invariant under the ${\bf Z}_2$ transformation.
This invariance is achieved as long as $g(y)=g(-y)$ and $h(y)=-h(-y)$.
We take the $g(y)$ to be $y$-independent and the $h(y)$ as
\begin{equation}
 h(y)=c {y \o |y|},
\end{equation}
where c is a constant to be determined in the next section.
The gauge symmetries are left unbroken with
the infinitesimal SU(3)$_C$ gauge transformation parameter 
restricted to satisfy $\c(x,y)=\c(x,-y)$.

\section{Boundary extra-quarks}

There are two fixed points in the $S^1/{\bf Z}_2$ orbifold: 
$y=0$ and $y=l$.
Let us put chiral extra-quarks\footnote{
Instead of boundary extra-quarks, we may introduce vector-like
extra-quarks in the bulk with a domain wall background,
which results in a similar separation of chiral extra-quarks
along the extra dimension.}
on the fixed-point boundaries:
\footnote{
The standard-model quarks may be put on one of the fixed points,
which does not alter the point in the following discussion.}
a left-handed extra-quark $\psi_L$ at $y=0$
and a right-handed extra-quark $\psi_R$ at $y=l$.
The action of the extra-quarks contains
\begin{equation}
 S_{\p} = \int_{y=0} \! d^4x \ {\bar \p_L}i{D\llap{/\,}}\p_L
        + \int_{y=l} \! d^4x \ {\bar \p_R}i{D\llap{/\,}}\p_R.
\end{equation}

Under an infinitesimal gauge transformation, this fermionic sector
provides a gauge anomaly
\beq
 \d S_{eff} = {i \o 24\pi^2}\int_{y=0}
            \tr \left(\c d(AdA+{1 \o 2}A^3) \right)
            - {i \o 24\pi^2}\int_{y=l}
            \tr \left(\c d(AdA+{1 \o 2}A^3) \right)
\eeq
due to its chirality,
though the fermion content is vector-like
from a four-dimensional perspective.

On the other hand,
the bulk action yields
\begin{eqnarray}
 \d S_A &=& \int h(y) \tr \left( (d \c) d(AdA+{1 \o 2}A^3) \right)
 \nonumber
 \\   
        &=& -\int (d h(y)) \tr \left( \c d(AdA+{1 \o 2}A^3) \right)
\end{eqnarray}
under the gauge transformation.
Gauge anomaly cancellation with the fermionic sector implies
\begin{equation}
c = {i \o 48\pi^2}.
\end{equation}

\section{Anomalous quasi-symmetry}

The gauge-invariant theory is given by the total action
$S = S_A + S_\p$.
Then, there is an approximate axial U(1) symmetry given 
by eq.(\ref{eq:chiral-transf}).
Indeed, interactions between the chiral extra-quarks
(in the four-dimensional effective theory) such as
\begin{equation}
 M_* \psi_L \psi^{\dagger}_{R} + {\rm h.c.}
\label{eq:LR-breaking}
\end{equation}
are suppressed by $e^{-M_* l}/(M_* l)$ with $M_*$ as the cutoff scale 
at which new particles and interactions may arise
(see also the discussion in the final section).

The extra-quarks should be decoupled from the low-energy spectrum to
escape from detection. 
Thus, we introduce hypercolor gauge interactions in the bulk
to confine the extra-quarks at high energies.
Such new gauge interactions at the same time induce a chiral condensate 
$\vev{\psi_{L} \psi^{\dagger}_R}$, break down the axial U(1) symmetry
\cite{KC}
and provide a corresponding Nambu-Goldstone (NG) boson
called an axion
\cite{WW}.
Non-vanishing anomaly U(1)[SU(3)$_C$]$^2$ induces a potential
of the axion field.

Let us adopt SU(3)$_H$ as the hypercolor gauge group and assume that
the chiral fermions on each boundary transform as 
$\psi_{L}({\bf 3},{\bf 3}^*)$ and $\psi_{R}({\bf 3},{\bf 3}^*)$
under the SU(3)$_C \times $SU(3)$_H$ gauge group.
\footnote{
Extensions to larger gauge groups
and fermion representations are straightforward,
which are touched upon in the final section.}
The SU(3)$_H$ interactions are supposed to be strong and
confining at an intermediate scale $F_a$, and
a chiral condensate $\vev{\psi_{L} \psi_{R}^{\dagger}} \simeq F^3_a$
develops.
Note that the [SU(3)$_H$]$^3$ anomaly due to the fermionic sector
can also be canceled by bulk Chern-Simons terms
in a similar way as in the previous section.

SU(3)$_H$-charged particles are confined and what is left at low
energies are only massless NG bosons.
When one switches off the SU(3)$_C$ gauge interactions (to
concentrate on the strong dynamics of the SU(3)$_H$ interactions),
then there is U(3)$_L \times$U(3)$_R$ flavor symmetry that acts on $\psi_L$
and $\psi_{R}^\dagger$, where the SU(3) subgroup of the 
diagonal U(3) symmetry is actually gauged as SU(3)$_C$.
The flavor symmetry U(3)$_L \times$U(3)$_R$ is spontaneously broken down 
to the diagonal U(3) symmetry. The NG bosons
due to this chiral symmetry breaking transform as ${\bf 3} \times
{\bf 3}^* = {\bf adj.}+{\bf 1}$ under the SU(3)$_C$.
However, there is not U(3)$_L \times$U(3)$_R$ symmetry actually since 
the diagonal SU(3) is gauged as the SU(3)$_C$ gauge group, and hence 
the adjoint-part of the NG bosons acquires masses due to 
the SU(3)$_C$ radiative corrections. 
What remains massless is only the color-singlet NG boson,
which corresponds to the axial U(1) symmetry in
eq.(\ref{eq:chiral-transf}). 

The axial symmetry discussed above, however, also has 
U(1)$\,$[$\, \SU(3)_H \,$]$^2$ anomaly.
Therefore, the color-singlet NG boson
obtains a large mass and it cannot play a role
of the axion for the color SU(3)$_C$.
Hence we further introduce an additional pair of chiral fermions 
on each boundary:
$\chi_{L}({\bf 1},{\bf 3}^*)$ at $y=0$
and 
$\chi_{R}({\bf 1},{\bf 3}^*)$ at $y=l$.
The global symmetry is now U(4)$_{L} \times$ U(4)$_{R}$
if the SU(3)$_C$ gauge interactions are neglected.
The strong dynamics of the SU(3)$_H$ gauge interactions
leads to spontaneous breakdown of the chiral symmetry,
$\vev{\psi_L \psi_{R}^{\dagger}} \simeq F_a^3$ and 
$\vev{\chi_L \chi_{R}^{\dagger}} \simeq F_a^3$.
Two color-singlets
would remain massless if it were not for anomalies.
In this case, one of them does play a role of the axion
that makes the effective strong CP phase
to be sufficiently small.

\section{Discussion}

The accidental chiral symmetry discussed above is broken 
by effective operators of the chiral fermions.
The operators involving both `$\psi_L$ or $\chi_L$' and
`$\psi_R^\dagger$ or $\chi_R^\dagger$', such as
eq.(\ref{eq:LR-breaking}),
are highly suppressed.
\footnote{
$M_* l \simeq 150 + 3 \ln (F_a/10^{12} \GEV)$ is sufficient.}
However, the operators involving either `$\psi_L$ and $\chi_L$' or
`$\psi_R^\dagger$ and $\chi_R^\dagger$' are expected to be suppressed
only by powers of $1/M_*$.
Thus, such operators induce an additional potential of the axion, 
but this correction is not too large to render the Peccei-Quinn
mechanism ineffective:
Axial-symmetry breaking on each boundary may have coupling
coefficients of order one.
Such operators of the lowest mass dimension\footnote{
Discussion here assumes that the extra-quarks are not charged under 
the SU(2)$_{\rm L}$ or U(1)$_{\rm Y}$ gauge group of the standard model,
though we can also consider a model where they are charged under the
SU(2)$_{\rm L}$ or U(1)$_{\rm Y}$.} are given by
\begin{equation}
               \int_{y=0} d^4x \frac{1}{M_*^5} (\psi_L)^3 (\psi_L)^3 + 
               \int_{y=l} d^4x \frac{1}{M_*^5} 
                                (\psi_R^\dagger)^3 (\psi_R^\dagger)^3 
               + {\rm h.c.}
\label{eq:LL-breaking}
\end{equation}
Integration of heavy particles of masses of order 
$F_a$ induces an additional potential of the axion field $a$ as 
\begin{equation}
 V(a) \simeq \frac{F_a^{14}}{M_*^{10}}f(\frac{a}{F_a}),
\end{equation}
where $f(a/F_a)$ is a function of $a/F_a$, whose minimum is generically 
different from that of the potential induced exclusively by the QCD effects.
The resulting shift in the $\theta_{\rm eff}$ parameter of the QCD
is expected to be of order
\begin{equation}
 {\theta_{\rm eff}} \simeq \left(\frac{F_a}{M_*}\right)^{14}
                         \left(\frac{M_*}{\Lambda_{\rm QCD}}\right)^4
                    = 10^{-10} \left(\frac{F_a 10^5}{M_*}\right)^{14}
                      \left(\frac{M_*}{\Lambda_{\rm QCD}10^{15}}\right)^4.
\end{equation}
The present experimental bound
$\theta_{\rm eff} < 10^{-9}$
is satisfied, for instance,
with $M_* \simeq 10^{18} \GEV$ and $F_a \leq 10^{12} \GEV$,
which is phenomenologically viable.

We note that the axial-symmetry breaking operators
on each boundary can be made to have higher mass dimensions
if we adopt a larger hypercolor gauge group
instead of the SU$(3)_H$. Then, the $\theta_{\rm eff}$ will be more
suppressed.

\section*{acknowledgment}

T.W. would like to thank
the Japan Society for the Promotion of Science for financial support.
This work was partially supported by ``Priority Area: Supersymmetry and
Unified Theory of Elementary Particles (\# 707)'' (T.Y.).

\end{document}